\documentclass[%
 reprint,
 amsmath,amssymb,
 aps,prd,showkeys,
floatfix,
]{revtex4-2}

\usepackage{graphicx}% Include figure files
\usepackage{dcolumn}% Align table columns on decimal point
\usepackage{bm}% bold math
\usepackage{placeins}
\begin{document}

\title{Large-Scale Cosmic-Ray Anisotropy in the TeV Regime: Peak Observables and a Propagation-Model Interpretation}% Force line breaks with \\

\author{Hao-Kun Chen}
\affiliation{School of Physics, Hebei Normal University, No. 20, East Section of the South 2nd Ring Road, Yuhua District, Shijiazhuang 050024, Hebei Province, China}

\author{Xu-Lin Dong}
\affiliation{School of Physics, Hebei Normal University, No. 20, East Section of the South 2nd Ring Road, Yuhua District, Shijiazhuang 050024, Hebei Province, China}
\affiliation{State Key Laboratory of Particle Astrophysics, Institute of High Energy Physics, Chinese Academy of Sciences, 19B Yuquan Road, Shijingshan District, Beijing 100049, China}

\author{Qi-Ling Chen}
\affiliation{College of Physics, Sichuan University, Chengdu 610064, China}

\author{Shu-Wang Cui}
\email[Contact author: ]{cuisw@hebtu.edu.cn}
\affiliation{School of Physics, Hebei Normal University, No. 20, East Section of the South 2nd Ring Road, Yuhua District, Shijiazhuang 050024, Hebei Province, China}
\affiliation{TIANFU Cosmic Ray Research Center, No. 1500, Kezhi Road, Xinglong Street, Shuangliu District, Chengdu 610213, Sichuan Province, China}

\author{Yi-Qing Guo}
\email[Contact author: ]{guoyq@ihep.ac.cn}
\affiliation{State Key Laboratory of Particle Astrophysics, Institute of High Energy Physics, Chinese Academy of Sciences, 19B Yuquan Road, Shijingshan District, Beijing 100049, China}
\affiliation{TIANFU Cosmic Ray Research Center, No. 1500, Kezhi Road, Xinglong Street, Shuangliu District, Chengdu 610213, Sichuan Province, China}
\affiliation{University of Chinese Academy of Sciences, Beijing 100049, China}

\date{\today}% It is always \today, today,
\begin{abstract}
In recent years, cosmic-ray observations have entered a stage of high-precision measurements, but the phases of large-scale anisotropy reported by different experiments differ significantly, partly because the first harmonic cannot fully describe non-dipolar structures. In this work, we process the one-dimensional relative-intensity distributions from ARGO-YBJ, Tibet AS$\gamma$, and HAWC using a unified procedure, perform multi-harmonic fits, and reconstruct the distributions, with the global maximum of each distribution used to define the peak amplitude $A_{\max}$ and the peak right ascension $\alpha_{\max}$. The results show that, in the common energy range of 11--22~TeV covered by the three experiments, the range of the published first-harmonic phases is $22.6^\circ$, whereas the range of $\alpha_{\max}$ is reduced to $7.5^\circ$; $\alpha_{\max}$ evolves overall from $60^\circ$--$70^\circ$ at the low-energy end to $250^\circ$--$300^\circ$ at the high-energy end, while $A_{\max}$ exhibits a local minimum and $\alpha_{\max}$ changes rapidly in the phase-flip region. At the low-energy end, $\alpha_{\max}$ lies between the right-ascension projections of the direction toward the Geminga supernova remnant and the direction of the local interstellar magnetic field. Accordingly, we construct a spatially dependent propagation model that includes Galactic background sources, a Geminga-related nearby source, and anisotropic diffusion along the direction of the local interstellar magnetic field. Within a unified framework, the model reproduces the energy spectra of P, He, C, O, and Fe, as well as the energy evolution of $A_{\max}$ and $\alpha_{\max}$, providing a self-consistent explanation for the origin of large-scale anisotropy in the TeV to tens-of-TeV energy range.
\end{abstract}

\maketitle

%\tableofcontents

\section{Introduction}

The origin and propagation of Galactic cosmic rays remain unresolved~\cite{Strong2007,Blasi2013}. Experiments including AMS-02, DAMPE, CALET, and LHAASO have identified multiple structures in measurements of energy spectra and composition~\cite{Aguilar2021AMS,An2019DAMPEProton,Adriani2019CALETProton,Cao2024LHAASO}. For example, DAMPE observed a common softening of carbon, oxygen, and iron at around 15~TV~\cite{DAMPE2026CHOFe}. Energy spectra mainly constrain source injection and propagation processes~\cite{Adriani2011PAMELA,Aguilar2015Proton,Yoon2017CREAMIII,Atkin2017NUCLEON}. Large-scale anisotropy further reflects the spatial gradient of the cosmic-ray density and the diffusion properties near the Solar System~\cite{BlasiAmato2012,Ahlers2016}. Therefore, jointly studying these two observables helps provide a more complete understanding of Galactic cosmic-ray propagation.

In the TeV--PeV energy range, the cosmic-ray anisotropy amplitude is approximately $10^{-4}$--$10^{-3}$~\cite{Guillian2007SuperK,Abdo2009Milagro,Abbasi2010IceCube}. Its direction changes markedly with energy~\cite{Aglietta2009EASTOP,Aartsen2016IceCube}. Nearby sources can affect both the local energy spectrum and anisotropy~\cite{Liu2019Nearby,Qiao2019}. The Geminga supernova remnant is one of the candidate sources~\cite{Zhao2022,Zhao2023}. Meanwhile, the local interstellar magnetic field can alter the direction of cosmic-ray flow through anisotropic diffusion~\cite{Effenberger2012}. Measurements of the direction of the local interstellar magnetic field near the Solar System by IBEX provide important clues for explaining the observed anisotropy direction~\cite{Schwadron2014,Funsten2013}.

However, reliably extracting the direction itself from observations is challenging. Traditionally, large-scale anisotropy is described by the first-harmonic amplitude $A_1$ and phase $\phi_1$ of the relative-intensity distribution in right ascension:
\begin{equation}
    I_1(\alpha)
    =
    1 + A_1 \cos\left(\alpha-\phi_1\right).
    \label{eq:first_harmonic}
\end{equation}

Actual distributions often contain non-sinusoidal structures such as the \textit{tail-in} and \textit{loss-cone}~\cite{Abeysekara2014HAWCSmall,Amenomori2017Tibet}. A first-harmonic fit determines the parameters by minimizing the sum of squared residuals over the entire distribution, and the cosmic-ray deficit region can cause $\phi_1$ to deviate from the location of the maximum in the excess region. Consequently, the first-harmonic phases published by different experiments may differ significantly at similar energies. This is related both to sky coverage and analysis methods and to the inadequacy of a single-dipole description~\cite{Bartoli2018ARGO,Abeysekara2018HAWC,GiacintiSigl2012,DesiatiLazarian2013}.

In this work, we select three Northern-sky experiments whose energy ranges overlap with one another while each also extends beyond the overlap: HAWC provides relatively dense measurement points from 2 to 72.8~TeV, ARGO-YBJ covers 4--520~TeV and spans the direction-transition region, and Tibet AS$\gamma$ extends the measurements to PeV energies. We apply a unified multi-harmonic fit to the one-dimensional relative-intensity distributions from the three experiments, define the peak amplitude $A_{\max}$ and peak right ascension $\alpha_{\max}$, and use them to re-examine the energy evolution of the anisotropy. We then construct a spatially dependent propagation model that includes a nearby Geminga source and anisotropic diffusion along the local interstellar magnetic field, and jointly interpret the energy spectra and the peak observables~\cite{Li2022Interpretation}.

\section{Data Processing and Definition of Peak Observables}
\subsection{Experimental Data and Unified Energy Binning}

We use the eight ARGO-YBJ energy bins (4, 7.1, 12, 22, 39, 71, 160, and 520~TeV)~\cite{Bartoli2018ARGO}, the five Tibet AS$\gamma$ energy bins (15, 50, 100, 300, and 1000~TeV)~\cite{Amenomori2017Tibet}, and the eight HAWC energy bins (2, 3, 4.4, 6.8, 11.2, 18.6, 30.3, and 72.8~TeV)~\cite{Abeysekara2018HAWC}. All three are Northern-sky experiments, and their observational fields of view have a high degree of overlap.

For ARGO-YBJ and Tibet AS$\gamma$, we use their respective publicly available one-dimensional relative-intensity distributions. Each energy bin contains 18 data points along right ascension, with adjacent points separated by $20^\circ$, and each point corresponds to the center of a $20^\circ$ interval. The published HAWC data are two-dimensional relative-intensity sky maps. In this work, the common color scale shared by all energy bins is used to construct a calibration table relating RGB colors to $\delta I$, which is then obtained through RGB matching. The relative intensity is defined as $I=1+\delta I$. The extracted values are first averaged over declination within each $2^\circ$ right-ascension interval. Ten consecutive $2^\circ$ right-ascension values are then averaged into a single $20^\circ$ bin, yielding the one-dimensional relative intensity at the center of each interval. In this way, the data from the three experiments have the same format and physical meaning, allowing a unified procedure to be used for the subsequent fitting. 

The right-ascension distribution is treated as a circular coordinate, with its beginning and end connected, so that a peak near $0^\circ$ is not split into two structures. For the low-energy region and the phase-flip region, where point-by-point uncertainties have not been publicly reported, we adopt a unified Poisson estimate. If the total number of events in a given energy bin is $N$, the expected number of events in each right-ascension interval is $N/18$, with a standard deviation of $\sqrt{N/18}$. The statistical uncertainty of the relative intensity is therefore
\begin{equation}
    \sigma_I
    =
    \frac{\sqrt{N/18}}{N/18}
    =
    \sqrt{\frac{18}{N}}.
    \label{eq:relative_intensity_uncertainty}
\end{equation}

The smaller the number of events, the larger the uncertainty. This estimate does not include experimental systematic uncertainties, digitization errors, or inter-bin correlations. In the high-energy region, the above estimate is compared with the uncertainties publicly reported by ARGO-YBJ and Tibet AS$\gamma$, and the two are generally consistent. The detailed procedure for extracting the one-dimensional relative-intensity distributions from the HAWC sky maps is described in Appendix~\ref{app:hawc_validation}.

\subsection{Multi-Harmonic Reconstruction and Peak Definition}

Considering that the deficit region may cause the first-harmonic phase to deviate from the peak direction of the excess region, we use a multi-harmonic function to fit and reconstruct the one-dimensional relative-intensity distribution:
\begin{equation}
    I_m(\alpha)
    =
    1
    +
    \sum_{n=1}^{m}
    A_n
    \cos\left[n\left(\alpha-\phi_n\right)\right],
    \qquad
    m=1,2,\ldots,7.
    \label{eq:multi_harmonic_reconstruction}
\end{equation}

An expansion up to order $m$ contains a total of $2m$ free parameters. Each distribution contains only 18 data points. For $m=7$, the fit contains 14 parameters and has 4 remaining degrees of freedom, whereas for $m=8$, only 2 degrees of freedom remain. To avoid overfitting, we therefore limit the maximum order to 7. After order selection, the selected orders for most energy bins are between 3 and 5. We determine the global maximum of the fitted function, define the deviation of this maximum from the mean intensity of 1 as the peak amplitude, and define the right ascension of the global maximum as the peak right ascension:
\begin{equation}
\begin{aligned}
    \alpha_{\max}
    &=
    \operatorname*{arg\,max} I_m(\alpha),
    \qquad
    0^\circ \leq \alpha < 360^\circ,
    \\
    A_{\max}
    &=
    I_m\left(\alpha_{\max}\right)-1.
\end{aligned}
\label{eq:peak_observables}
\end{equation}

For distributions that are close to a dipole, $A_1$ and $A_{\max}$ in the same energy bin are numerically similar. We focus on the difference between $\alpha_{\max}$ and the first-harmonic phase $\phi_1$. Here, $\phi_1$ is the first-harmonic phase, whereas $\alpha_{\max}$ is the right-ascension direction corresponding to the region of maximum excess in the multi-harmonic reconstructed distribution; the two quantities have different physical meanings.

\subsection{Order Selection and Stability Tests}
The primary criterion for order selection is whether the peak right ascension is stable. Specifically, starting from the highest order, $m=7$, we calculate the circular difference between the peak right ascensions of adjacent orders:
\begin{equation}
    \Delta\alpha_{m,m-1}
    =
    \min\left(
        \left|\alpha_m-\alpha_{m-1}\right|,
        360^\circ-\left|\alpha_m-\alpha_{m-1}\right|
    \right).
    \label{eq:circular_ra_difference}
\end{equation}

If $\Delta\alpha_{7,6}\leq 2^\circ$, the peak right ascensions obtained at orders 7 and 6 are considered to have converged, after which the test proceeds successively toward lower orders. We select the lowest order for which the differences in peak right ascension between adjacent orders do not exceed the threshold. This procedure is referred to as the ``convergence criterion.'' The threshold is set to one tenth of the $20^\circ$ spacing between adjacent data points. After relaxing the threshold from $2^\circ$ to $3^\circ$ and repeating the complete procedure, the resulting $A_{\max}$, $\alpha_{\max}$, and their overall evolution show no substantive changes, indicating that the conclusions are insensitive to this threshold.

For energy bins that do not satisfy the above conditions, we use the AIC to select the order~\cite{Akaike1974}:
\begin{equation}
    \mathrm{AIC}_m
    =
    \chi_m^2+2k_m
    =
    \chi_m^2+4m.
    \label{eq:aic}
\end{equation}

Here, $k_m=2m$ is the number of fitting parameters. The AIC achieves a balance between goodness of fit and model complexity, and the order with the minimum AIC is selected. The three experiments contain a total of 21 energy bins, of which 14 satisfy the convergence criterion: 4, 7.1, 12, 22, 39, and 520~TeV from ARGO-YBJ; 15 and 100~TeV from Tibet AS$\gamma$; and 2, 3, 4.4, 6.8, 11.2, and 18.6~TeV from HAWC. The remaining seven energy bins use the AIC for order selection, thereby yielding more appropriate orders. Each candidate order $m$ is fitted separately, and the corresponding $\chi_m^2$ is calculated. The first term in the formula, $\chi_m^2$, is the goodness-of-fit term, for which a smaller value is better; the second term, $2k_m=4m$, is the complexity-penalty term, for which the penalty becomes stronger as $m$ increases. Therefore, the order that minimizes the AIC is selected.

Taking the 22~TeV ARGO-YBJ energy bin as an example, the peak right ascensions at $m=5$, 6, and 7 are $69.9^\circ$, $71.3^\circ$, and $69.7^\circ$, respectively. The differences between adjacent higher orders are all within $2^\circ$, whereas the difference between orders 5 and 4 is $4.1^\circ$, exceeding the threshold. We therefore select $m=5$, corresponding to $\alpha_{\max}=69.9^\circ$. This value differs from the originally published first-harmonic phase of $22.3^\circ$, indicating that the distribution is not dipolar and that peak right ascension and dipole phase have different meanings. To illustrate this difference, Appendix~\ref{app:argo22_harmonic_fits} compares the first-harmonic and multi-harmonic fits to the 22~TeV ARGO-YBJ relative-intensity distribution.

To evaluate the uncertainties associated with order selection and peak extraction, we perform $10^4$ Monte Carlo tests for each energy bin. For each data point in the publicly available distribution, we draw samples using its measured value as the mean and its estimated uncertainty as the standard deviation, generate simulated distributions, and repeat the complete multi-harmonic fitting, order-selection, and peak-extraction procedure. In the low-energy region, the $\alpha_{\max}$ distribution is concentrated and approximately unimodal. In the phase-flip region, the selected order may differ from that obtained from the actual data, suggesting that the corresponding energy bin may contain two nearby local maxima or a flat maximum, causing the $\alpha_{\max}$ distribution to become multimodal or significantly broadened. In the high-energy region, the distribution remains unimodal but is more broadly distributed. Repeating the complete order-selection procedure incorporates variations in the selected order into the uncertainty and avoids underestimating the variation in $\alpha_{\max}$ by fitting only at a fixed order.

\subsection{Cross-Experiment Consistency and Energy Evolution}

\begin{figure*}[!t]
    \centering
    \includegraphics[width=0.85\textwidth]{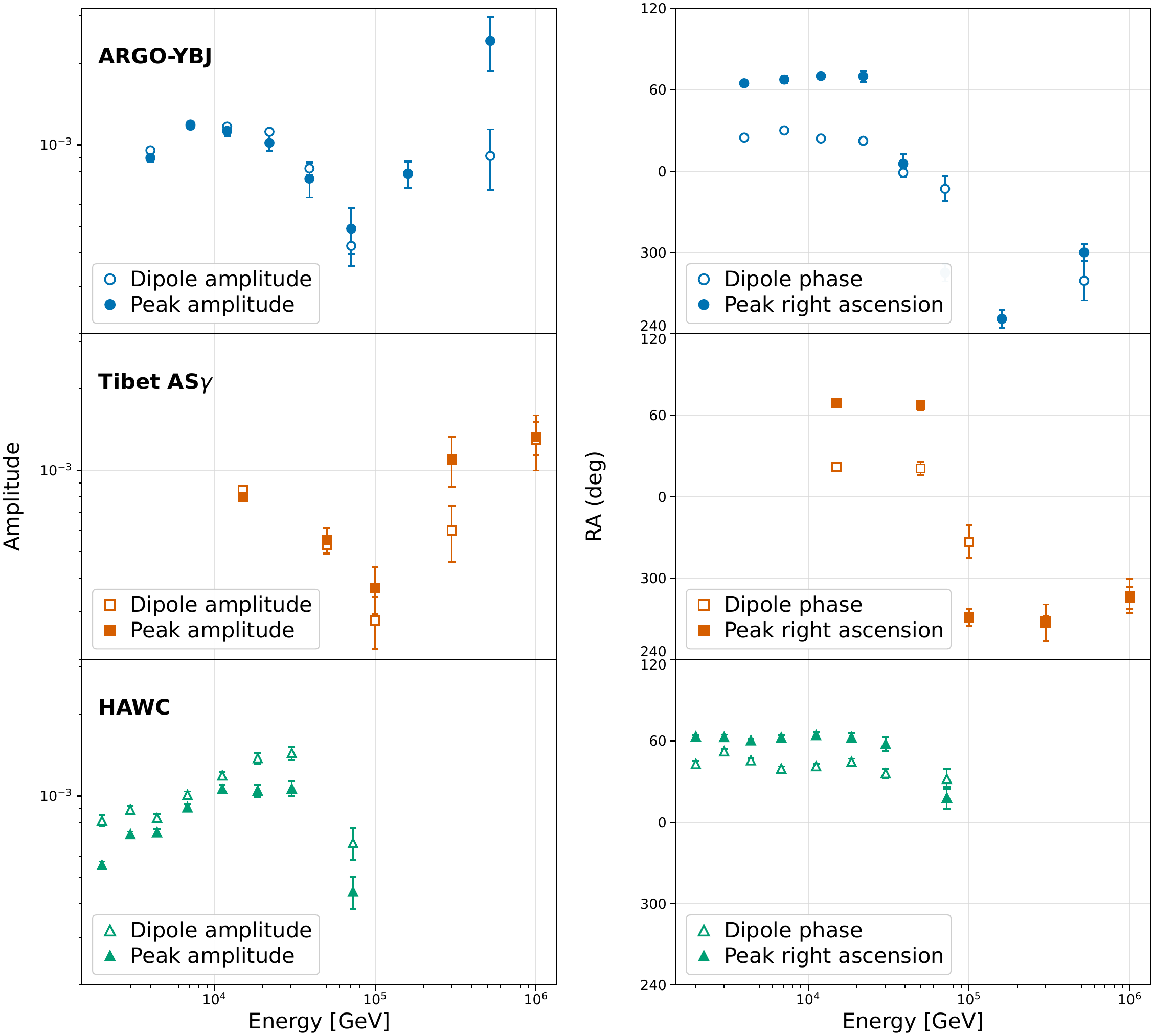}
    \caption{
        Comparison of the published dipole amplitudes and dipole phases
        with the peak amplitudes and peak right ascensions for the three
        experiments. Open symbols denote $A_1$ and $\phi_1$, whereas
        filled symbols denote $A_{\max}$ and $\alpha_{\max}$. From top
        to bottom, the panels correspond to ARGO-YBJ, Tibet AS$\gamma$,
        and HAWC, respectively. The experimental data are taken from
        Refs.~\cite{Bartoli2018ARGO,Amenomori2017Tibet,Abeysekara2018HAWC}.
    }
    \label{fig:published_and_peak_observables}
\end{figure*}

\begin{figure*}[!t]
    \centering
    \includegraphics[width=0.85\textwidth]{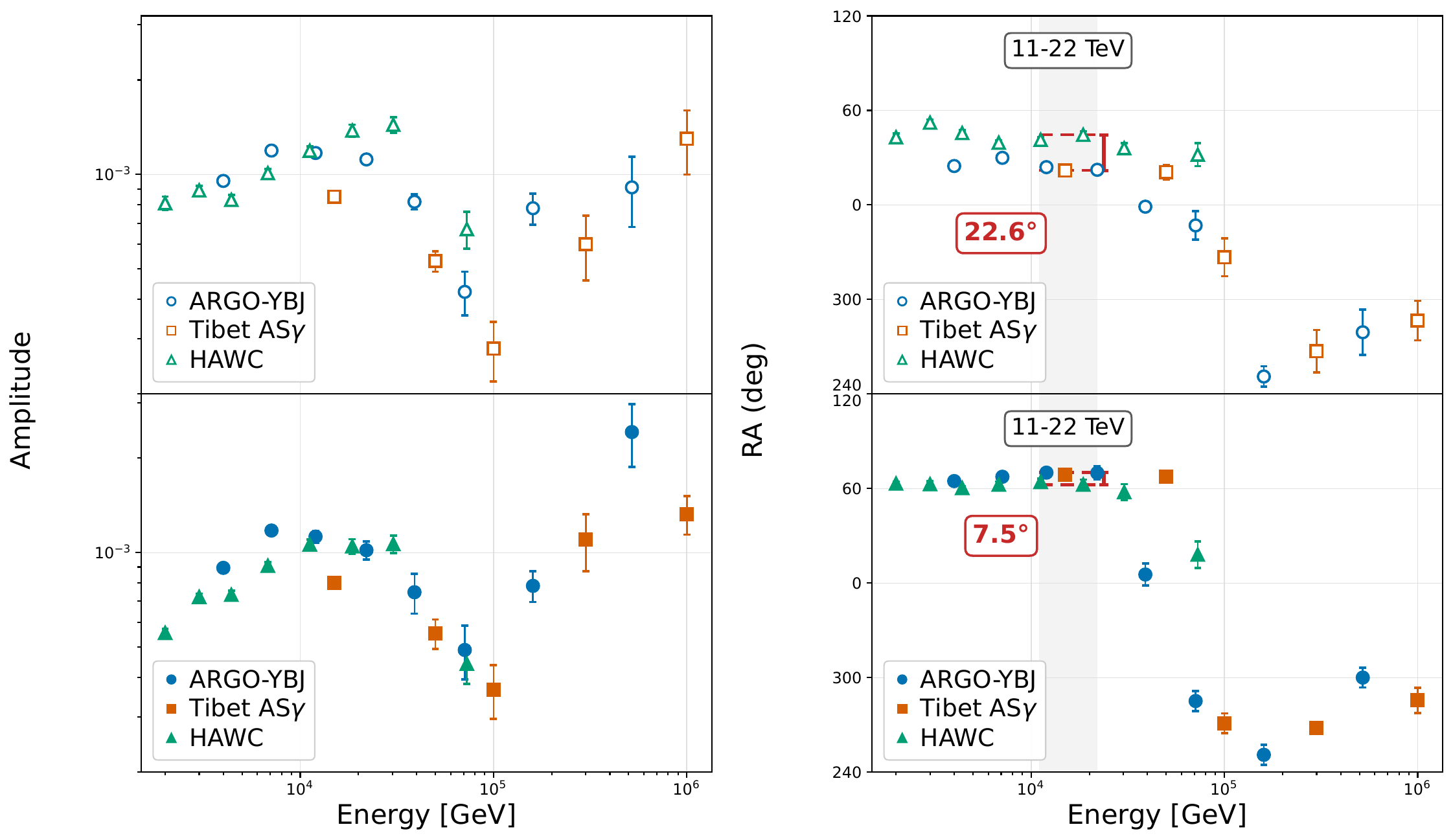}
    \caption{
        Combined comparison of the three experiments. The upper-left
        panel shows $A_1$, the upper-right panel shows $\phi_1$, the
        lower-left panel shows $A_{\max}$, and the lower-right panel
        shows $\alpha_{\max}$. The red boxes indicate the circular
        ranges of the phase and peak right ascension among the three
        experiments in the 11--22~TeV energy range. The ARGO-YBJ,
        Tibet AS$\gamma$, and HAWC data are taken from
        Refs.~\cite{Bartoli2018ARGO,Amenomori2017Tibet,Abeysekara2018HAWC}.
    }
    \label{fig:three_experiment_summary}
\end{figure*}

Figure~\ref{fig:published_and_peak_observables} compares $A_1$ and $\phi_1$ with $A_{\max}$ and $\alpha_{\max}$ for each energy bin of the three experiments, while Fig.~\ref{fig:three_experiment_summary} presents the combined results of the three experiments.

The main results are as follows. First, in the common energy range of 11--22~TeV covered by the three experiments, the published first-harmonic phases span $21.9^\circ$--$44.5^\circ$, corresponding to a range of $22.6^\circ$, whereas $\alpha_{\max}$ spans $62.58^\circ$--$70.11^\circ$ in this energy range, with a range of only $7.5^\circ$. After adopting the peak right ascension, the consistency among the experiments is significantly improved. Second, the values of $\alpha_{\max}$ in the low-energy regions of all three experiments are concentrated within $60^\circ$--$70^\circ$, and this interval lies between the right-ascension projections of the direction toward the Geminga supernova remnant and the IBEX local interstellar magnetic-field direction. Third, as the energy increases, $\alpha_{\max}$ evolves overall from $60^\circ$--$70^\circ$ at the low-energy end to $250^\circ$--$300^\circ$ at the high-energy end; in the phase-flip region, $A_{\max}$ exhibits a local minimum, while $\alpha_{\max}$ changes rapidly. At high energies, $A_{\max}$ increases again at some data points, but because of the influence of statistical uncertainties, the trend should be assessed primarily from the overall evolution.

The above geometrical relationship suggests that, at the low-energy end, the cosmic-ray excess direction may result from the combined effects of the local density gradient produced by a nearby Geminga source and anisotropic diffusion in the local interstellar magnetic field. In the following, we construct a propagation model to test this interpretation.

\section{Propagation Model and Joint Interpretation}
\subsection{Spatially Dependent Propagation Framework}
The spatially dependent propagation model divides the diffusion region into an inner halo and an outer halo~\cite{GuoYuan2018,Guo2016,Liu2018SDP}. The diffusion properties in the outer halo are the same as those in the conventional model. In the inner halo, diffusion is affected by the source distribution: source-dense regions have stronger turbulence and slower diffusion, whereas source-sparse regions have faster diffusion. The magnetic halo is approximated as a cylinder, with the Galactic disk located at its center and a radius of approximately 20~kpc. Sources and the interstellar medium are distributed mainly within the Galactic disk, and the vertical scale of the source distribution is approximately 100~pc. The transport equation and boundary conditions are the same as those in the conventional model~\cite{StrongMoskalenko1998,Evoli2008DRAGON,Evoli2017}. Under the steady-state condition, the transport equation can be written as
\begin{equation}
\begin{aligned}
\frac{\partial\Psi(\boldsymbol{r},p,t)}{\partial t}
={}&
Q(\boldsymbol{r},p,t)
\\
&+
\boldsymbol{\nabla}\cdot
\left(
D_{xx}\boldsymbol{\nabla}\Psi
-\boldsymbol{v}_{c}\Psi
\right)
\\
&+
\frac{\partial}{\partial p}
\left[
p^2D_{pp}
\frac{\partial}{\partial p}
\left(
\frac{\Psi}{p^2}
\right)
\right]
\\
&-
\frac{\partial}{\partial p}
\left[
\dot{p}\Psi
-\frac{p}{3}
\left(
\boldsymbol{\nabla}\cdot\boldsymbol{v}_{c}
\right)\Psi
\right]
\\
&-
\frac{\Psi}{\tau_f}
-
\frac{\Psi}{\tau_r}.
\end{aligned}
\label{eq:transport_equation}
\end{equation}

Here, $\Psi(\boldsymbol{r},p,t)$ denotes the cosmic-ray density in phase space, and $Q(\boldsymbol{r},p,t)$ denotes the source-injection term. The remaining terms describe cosmic-ray diffusion and loss processes: $D_{xx}$ describes spatial diffusion, $\boldsymbol{v}_{c}$ denotes the convection velocity, the momentum-diffusion coefficient $D_{pp}$ describes diffusive reacceleration~\cite{Seo1994}, $\dot{p}$ denotes continuous energy loss, and $\tau_f$ and $\tau_r$ are the characteristic timescales of fragmentation and radioactive decay, respectively. Under the steady-state condition, the source term is decomposed into a spatial distribution and an injection spectrum:
\begin{equation}
    Q(\boldsymbol{r},p)
    =
    f(r,z)q(R).
    \label{eq:source_term}
\end{equation}

Here, $f(r,z)$ represents the spatial distribution of background sources in the Galaxy, and $q(R)$ describes the rigidity dependence of the injection spectrum. The spatial distribution of the background sources is described using the commonly adopted disk-like parameterization~\cite{Green2015}:
\begin{equation}
    f(r,z)
    =
    \left(
        \frac{r}{r_{\odot}}
    \right)^{1.09}
    \exp\left[
        -\frac{3.87\left(r-r_{\odot}\right)}{r_{\odot}}
    \right]
    \exp\left(
        -\frac{|z|}{z_s}
    \right).
    \label{eq:background_source_distribution}
\end{equation}

In this work, we adopt $r_{\odot}=8.5~\mathrm{kpc}$ and $z_s=0.1~\mathrm{kpc}$. The injection spectrum is a rigidity-dependent broken power-law spectrum:
\begin{equation}
q(R)
=
q_0
\begin{cases}
\displaystyle
\left(
    \frac{R}{R_{\mathrm{br}}}
\right)^{-\nu_1},
&
R\leq R_{\mathrm{br}},
\\[8pt]
\displaystyle
\left(
    \frac{R}{R_{\mathrm{br}}}
\right)^{-\nu_2}
\exp\left(
    -\frac{R}{R_c}
\right),
&
R>R_{\mathrm{br}}.
\end{cases}
\label{eq:injection_spectrum}
\end{equation}

Here, $R_{\mathrm{br}}$ is the break rigidity, $R_c$ is the cutoff rigidity, and $\nu_1$ and $\nu_2$ are the injection-spectrum indices before and after the break, respectively. The diffusion coefficient is written as~\cite{Dong2026SpectralSoftenings}
\begin{equation}
    D_{xx}(r,z,R)
    =
    D_0 F(r,z)\,
    \beta^{\eta}
    \left(
        \frac{R}{R_0}
    \right)^{\delta_0F(r,z)}.
    \label{eq:diffusion_coefficient}
\end{equation}

Here, $D_0$ is the normalization at the reference rigidity $R_0=4~\mathrm{GV}$, $\beta=v/c$, and $\delta_0$ controls the rigidity dependence. The factor $\beta^\eta$ is used to describe the low-energy correction. In this study, we adopt $\beta\simeq1$. The spatial modulation function $F(r,z)$ is controlled by Eq.~(\ref{eq:spatial_modulation}):
\begin{equation}
F(r,z)
=
\begin{cases}
\displaystyle
g(r,z)
+
\left[
    1-g(r,z)
\right]
\left(
    \frac{|z|}{\xi z_0}
\right)^n,
&
|z|\leq \xi z_0,
\\[8pt]
1,
&
|z|>\xi z_0.
\end{cases}
\label{eq:spatial_modulation}
\end{equation}

Here,
\begin{equation}
    g(r,z)
    =
    \frac{N_m}{1+f(r,z)}.
    \label{eq:diffusion_suppression}
\end{equation}

In this expression, $N_m$ is the normalization factor that controls the degree of diffusion suppression, $\xi$ is the ratio of the inner-halo thickness to the total halo thickness, and $n$ controls the smoothness of the transition between the inner and outer halos. The main parameters are listed in Table~\ref{tab:propagation_parameters}.

\begin{table}[!t]
    \caption{Main parameters of the spatially dependent propagation model.}
    \label{tab:propagation_parameters}
    \centering

    \renewcommand{\arraystretch}{1.35}

    \begin{minipage}{0.78\columnwidth}
    \centering
    \begin{ruledtabular}
    \begin{tabular}{ccc}
        Parameter & Value & Unit \\
        \colrule
        $D_0$      & $4.87\times10^{28}$ & $\mathrm{cm^2\,s^{-1}}$ \\
        $N_m$      & $0.62$               & \\
        $\delta_0$ & $0.58$               & \\
        $R_0$      & $4$                  & $\mathrm{GV}$ \\
        $\xi$      & $0.10$               & \\
        $z_s$      & $0.1$                & $\mathrm{kpc}$ \\
        $v_A$      & $6$                  & $\mathrm{km\,s^{-1}}$ \\
        $z_0$      & $5$                  & $\mathrm{kpc}$ \\
        $n$        & $4$                  & \\
    \end{tabular}
    \end{ruledtabular}
    \end{minipage}
\end{table}

\subsection{Geminga-Related Source and Energy-Spectrum Constraints}

To characterize both the energy spectra and the anisotropy, we treat the Geminga supernova remnant in the Galactic anticenter direction as a nearby source and describe it as a single instantaneous injection~\cite{Dong2026SpectralSoftenings}:
\begin{equation}
\begin{aligned}
Q_{\mathrm{Gem}}(R,t)
={}&
q_{\mathrm{Gem}}\,
\delta\left(t-t_s\right)
\left(
\frac{R}{R_{\mathrm{ref}}}
\right)^{-\gamma_{\mathrm{Gem}}}
\\
&\times
\exp\left(
-\frac{R}{R_{\mathrm{cut,Gem}}}
\right).
\end{aligned}
\label{eq:geminga_injection}
\end{equation}

\begin{figure*}[!t]
    \centering
    \includegraphics[width=0.85\textwidth]{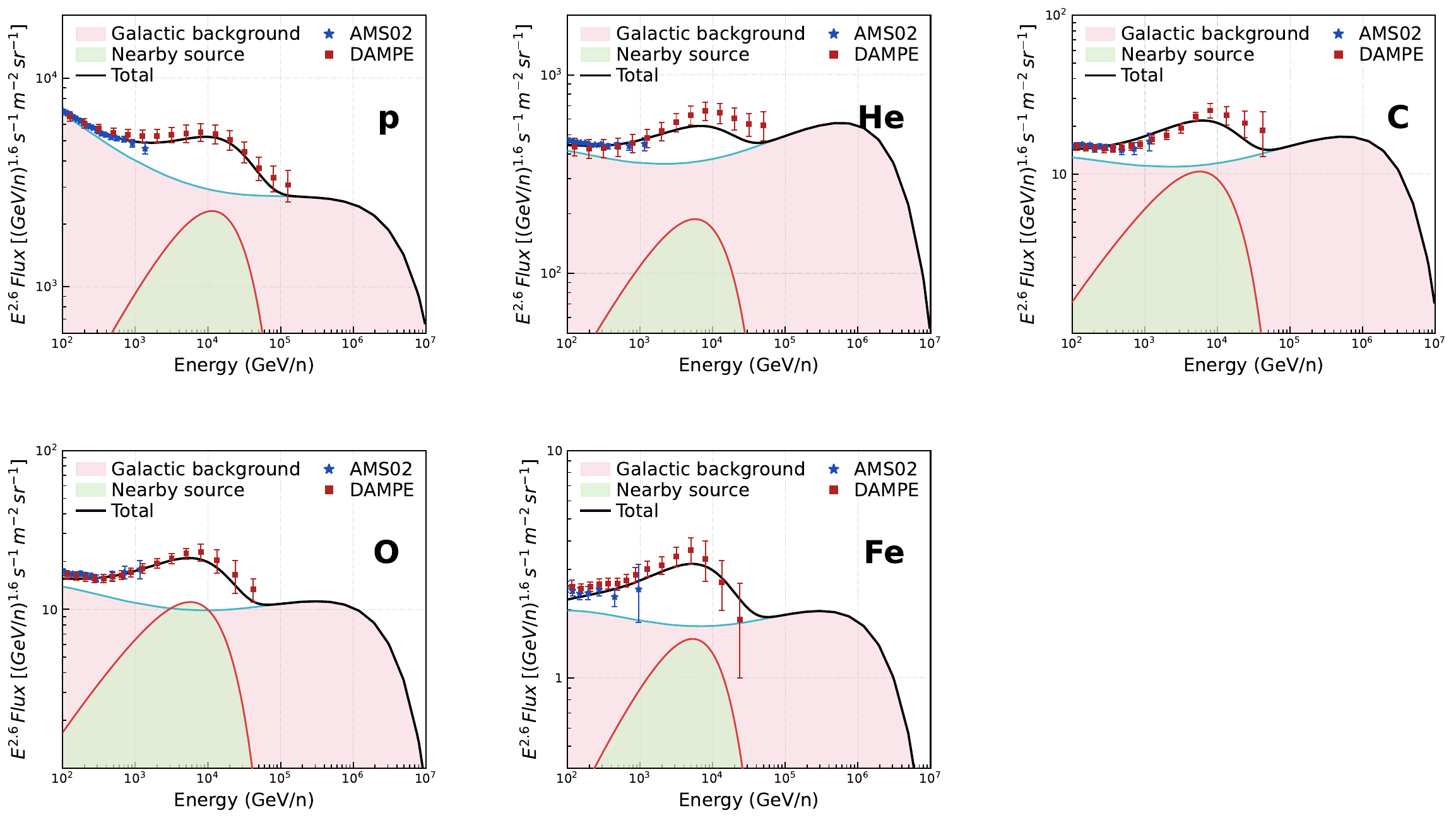}
    \caption{
        Energy spectra of $P$, He, C, O, and Fe. The pink component
        represents the contribution from Galactic background sources,
        the light-green component represents the contribution from the
        Geminga-related nearby source, and the solid black line is the
        sum of the two components. The blue data points are from
        AMS-02~\cite{Aguilar2021AMS}, and the red data points are from
        DAMPE~\cite{An2019DAMPEProton,Alemanno2021DAMPEHelium,DAMPE2026CHOFe}.
    }
    \label{fig:nuclear_energy_spectra}
\end{figure*}

Here, $t_s$ denotes the time at which the remnant underwent the instantaneous injection; its propagation age is defined as $t_{\mathrm{Gem}}=t-t_s$, $q_{\mathrm{Gem}}$ is the normalization coefficient, $\gamma_{\mathrm{Gem}}$ is the injection-spectrum index, and $R_{\mathrm{cut,Gem}}$ is the cutoff rigidity. The position, age, and distance of the Geminga supernova remnant are given in Table~\ref{tab:geminga_injection_parameters} and its table note. The normalization coefficients are determined by fitting the $P$, He, C, O, and Fe energy spectra measured by AMS-02 and DAMPE. The results are shown in Fig.~\ref{fig:nuclear_energy_spectra}, and the detailed injection parameters are listed in Table~\ref{tab:geminga_injection_parameters}.

\begin{table*}[!t]
    \caption{
        Galactic background-flux parameters and injection parameters
        of the nearby Geminga source.
    }
    \label{tab:geminga_injection_parameters}
    \centering

    {%
    \renewcommand{\arraystretch}{1.40} % 行间距
    \setlength{\tabcolsep}{4pt}        % 单元格左右留白

    \begin{minipage}{0.85\textwidth}   % 控制表格整体宽度
    \centering
    \begin{ruledtabular}
    \begin{tabular}{ccccc}
        Nucleus
        &
        \begin{tabular}{c}
        Background-source\\
        normalization\\
        $\left(\mathrm{GeV}/n\right)^{-1}
        \mathrm{m^{-2}\,s^{-1}\,sr^{-1}}$
        \end{tabular}
        &
        \begin{tabular}{c}
        Background-source\\
        spectral index $\nu_2$
        \end{tabular}
        &
        \begin{tabular}{c}
        Geminga\\
        normalization\\
        $\left(\mathrm{GeV}^{-1}\right)$
        \end{tabular}
        &
        \begin{tabular}{c}
        Geminga\\
        spectral index
        \end{tabular}
        \\
        \colrule
        $p$ & $4.26\times10^{-2}$ & $2.373$
            & $4.68\times10^{51}$ & $1.83$ \\
        He  & $2.32\times10^{-3}$ & $2.295$
            & $2.3088\times10^{51}$ & $1.83$ \\
        C   & $9.19\times10^{-5}$ & $2.285$
            & $1.2784\times10^{50}$ & $1.83$ \\
        O   & $1.20\times10^{-4}$ & $2.334$
            & $1.3693\times10^{50}$ & $1.83$ \\
        Fe  & $2.12\times10^{-5}$ & $2.428$
            & $2.2127\times10^{49}$ & $1.83$ \\
    \end{tabular}
    \end{ruledtabular}
    \end{minipage}
    }%

    \vspace{2mm}

    \begin{minipage}{0.96\textwidth}
    \footnotesize
    The background-source normalization is the differential flux at
    $100~\mathrm{GeV}/n$. The remaining parameters are
    $d_{\mathrm{Gem}}=0.33~\mathrm{kpc}$,
    $t_{\mathrm{Gem}}=3.4\times10^{5}~\mathrm{yr}$,
    $(l_{\mathrm{Gem}},b_{\mathrm{Gem}})
    =(194.3^\circ,-13.0^\circ)$,
    $R_{\mathrm{cut,Gem}}=20~\mathrm{TV}$, and
    $R_{\mathrm{ref}}=1~\mathrm{GV}$.
    \end{minipage}
\end{table*}

\subsection{Local Interstellar Magnetic Field and Anisotropy Vector}

Within the diffusion picture, large-scale anisotropy originates from the nonuniformity of the spatial distribution of cosmic rays near the Solar System. The Galactic source distribution, spatially dependent diffusion, and the contribution from a nearby source can all produce a nonzero density gradient near the Solar System, thereby driving a net diffusive flux. Adopting a scalar-diffusion approximation first, the dipole-anisotropy vector can be estimated from the local density gradient as~\cite{Dong2026SpectralSoftenings}
\begin{equation}
    \boldsymbol{\Delta}(R)
    =
    \left.
    \frac{
        3D_{xx}\left(R,\boldsymbol{r}_{\odot}\right)
    }{
        v\Psi\left(R,\boldsymbol{r}_{\odot}\right)
    }
    \boldsymbol{\nabla}\Psi\left(R,\boldsymbol{r}\right)
    \right|_{\boldsymbol{r}=\boldsymbol{r}_{\odot}}.
    \label{eq:scalar_anisotropy}
\end{equation}

Here, $v$ is the particle velocity, $\Psi(R,\boldsymbol{r}_{\odot})$ is the cosmic-ray density at the position of the Solar System, and $\boldsymbol{\nabla}\Psi$ is the local spatial-density gradient. This expression shows that the anisotropy amplitude is jointly determined by the diffusion coefficient and the relative density gradient. The local interstellar magnetic field causes the diffusion efficiencies parallel and perpendicular to the magnetic-field direction to differ~\cite{Jokipii1966,GiacaloneJokipii1999,Casse2001}. We therefore further adopt the diffusion tensor
\begin{equation}
    D_{ij}
    =
    D_{\perp}\delta_{ij}
    +
    \left(
        D_{\parallel}-D_{\perp}
    \right)b_i b_j.
    \label{eq:diffusion_tensor}
\end{equation}

Here, $\delta_{ij}$ is the identity tensor, and
$\boldsymbol{b}=\boldsymbol{B}_0/|\boldsymbol{B}_0|$ is the unit vector
along the direction of the local regular magnetic field.
$D_{\parallel}$ and $D_{\perp}$ denote the parallel and perpendicular
diffusion coefficients, respectively. When
$D_{\parallel}>D_{\perp}$, parallel diffusion is more efficient, and
the diffusive flux under the action of the tensor tends to be directed
more closely along the magnetic field. However, its final direction is
still jointly determined by the angle between the density gradient and
the magnetic field.

Following the existing treatment, we adopt
\begin{equation}
    D_{\parallel}(R)
    =
    D_{xx}\left(r_{\odot},0,R\right),
    \label{eq:parallel_diffusion}
\end{equation}
and set the relation between the perpendicular and parallel diffusion
coefficients to
\begin{equation}
    \frac{D_{\perp}(R)}{D_{\parallel}(R)}
    =
    \min\left[
        \epsilon
        \left(
            \frac{R}{R_0}
        \right)^{\zeta},
        1
    \right].
    \label{eq:perpendicular_parallel_ratio}
\end{equation}

Here, $\epsilon$ is the ratio of perpendicular diffusion to parallel
diffusion at the reference rigidity, and $\zeta$ describes the
rigidity dependence of this ratio. The upper limit ensures that
$D_{\perp}\leq D_{\parallel}$; when the ratio reaches unity, local
diffusion becomes approximately isotropic. We adopt the IBEX local
interstellar magnetic-field direction
\[
    \left(l_{\mathrm{IBEX}},b_{\mathrm{IBEX}}\right)
    =
    \left(210.5^\circ,-57.1^\circ\right),
\]
whose projection in right ascension is approximately $48.5^\circ$.
The anisotropic-diffusion parameters are listed in
Table~\ref{tab:anisotropic_diffusion_parameters}. In the calculations,
we test two values, $\epsilon=0.10$ and $0.15$, while keeping all other
parameters unchanged.

\begin{table*}[!t]
    \caption{
        Local interstellar magnetic-field and anisotropic-diffusion
        parameters.
    }
    \label{tab:anisotropic_diffusion_parameters}
    \centering

    {%
    \renewcommand{\arraystretch}{1.5} % 行间距
    \setlength{\tabcolsep}{4pt}        % 单元格左右留白

    \begin{minipage}{0.75\textwidth}   % 控制表格整体宽度
    \centering
    \begin{ruledtabular}
    \begin{tabular}{ccccccc}
        Parameter
        & $l_{\mathrm{IBEX}}$
        & $b_{\mathrm{IBEX}}$
        & $R_0$
        & $\zeta$
        & $\epsilon$
        & $D_{\parallel}$
        \\
        \colrule
        Value
        & $210.5^\circ$
        & $-57.1^\circ$
        & $4~\mathrm{GV}$
        & $0.20$
        & $0.10,\ 0.15$
        & $D_{xx}$
        \\
    \end{tabular}
    \end{ruledtabular}
    \end{minipage}

    }%
\end{table*}

After introducing the anisotropic-diffusion tensor, the anisotropy
vector of the $i$th nuclear species is written as
\begin{equation}
    \boldsymbol{\Delta}_i(E)
    =
    \left.
    \frac{
        3\mathbf{D}_i\left(E,\boldsymbol{r}_{\odot}\right)
    }{
        v_i\Psi_i\left(E,\boldsymbol{r}_{\odot}\right)
    }
    \cdot
    \boldsymbol{\nabla}
    \Psi_i\left(E,\boldsymbol{r}\right)
    \right|_{\boldsymbol{r}=\boldsymbol{r}_{\odot}}.
    \label{eq:species_anisotropy_vector}
\end{equation}

This expression distinguishes the different roles of the source distribution and the local interstellar magnetic field. The Galactic background sources and the Geminga supernova remnant jointly determine
the magnitude and direction of the density gradient
$\boldsymbol{\nabla}\Psi$ near the Solar System, whereas the local interstellar magnetic field changes both the direction and magnitude of the diffusive flux corresponding to this gradient through the diffusion tensor $\mathbf{D}_i$. Therefore, the anisotropy direction observed experimentally does not necessarily coincide exactly with the direction of Geminga, the direction of the Galactic center, or the direction of the density gradient~\cite{Li2024Local}.

The total anisotropy vector is obtained by a flux-weighted sum:
\begin{equation}
    \boldsymbol{\Delta}_{\mathrm{tot}}(E)
    =
    \frac{
        \displaystyle
        \sum_i
        \Phi_i(E)\boldsymbol{\Delta}_i(E)
    }{
        \displaystyle
        \sum_i\Phi_i(E)
    }.
    \label{eq:total_anisotropy_vector}
\end{equation}

The amplitude is $A(E)=|\boldsymbol{\Delta}_{\mathrm{tot}}(E)|$, and the phase is determined by the direction of $\boldsymbol{\Delta}_{\mathrm{tot}}$. The anisotropies of different nuclear species may reinforce or partially cancel one another; therefore, vector summation must be performed.

\subsection{Model Predictions for $A_{\max}$ and $\alpha_{\max}$}

The model relative-intensity distribution is written as
\begin{equation}
    I_{\mathrm{mod}}(\alpha,E)
    =
    1
    +
    A(E)
    \cos\left[
        \alpha-\phi_{\mathrm{tot}}(E)
    \right].
    \label{eq:model_relative_intensity}
\end{equation}
Here, $A(E)$ is the model anisotropy amplitude, and $\phi_{\mathrm{tot}}(E)$ is the phase determined by the right-ascension projection of $\boldsymbol{\Delta}_{\mathrm{tot}}(E)$. We then adopt the same maximum-based definitions as those used for the experimental data:
\begin{equation}
\begin{aligned}
    \alpha_{\max}^{\mathrm{mod}}(E)
    &=
    \operatorname*{arg\,max}
    I_{\mathrm{mod}}(\alpha,E),
    \qquad
    0^\circ\leq\alpha<360^\circ,
    \\
    A_{\max}^{\mathrm{mod}}(E)
    &=
    I_{\mathrm{mod}}
    \left(
        \alpha_{\max}^{\mathrm{mod}}(E),E
    \right)-1.
\end{aligned}
\label{eq:model_peak_observables}
\end{equation}

If the model gives a dipole approximation,
$\alpha_{\max}^{\mathrm{mod}}$ is the vector direction, allowing the
directions of maximum relative intensity obtained from the model and
the experiments to be compared.

\begin{figure*}[!t]
    \centering
    \includegraphics[width=0.75\textwidth]{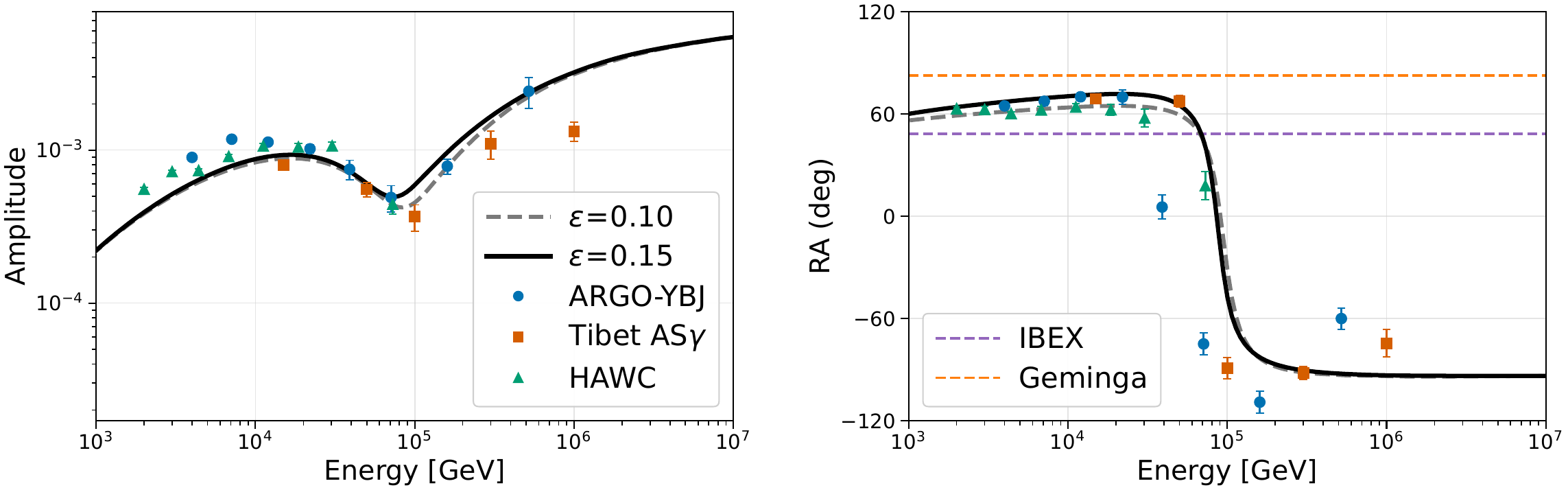}
    \caption{
        Comparison of the peak amplitude and peak right ascension with
        the model calculations. The left panel shows the peak amplitude
        $A_{\max}$, and the right panel shows the peak right ascension
        $\alpha_{\max}$. The gray dashed and black solid curves
        correspond to $\epsilon=0.10$ and $0.15$, respectively. The
        blue circles, red squares, and green triangles represent
        ARGO-YBJ, Tibet-AS$\gamma$, and HAWC, respectively, with the
        experimental data taken from
        Refs.~\cite{Bartoli2018ARGO,Amenomori2017Tibet,Abeysekara2018HAWC}.
        In the
        right panel, the orange horizontal dashed line indicates the
        projection of Geminga in right ascension, approximately
        $82.5^\circ$~\cite{Zhao2022}, and the purple horizontal dashed line indicates
        the projection of the IBEX local interstellar magnetic field
        in right ascension, approximately
        $48.5^\circ$~\cite{Schwadron2014,Funsten2013}.
    }
    \label{fig:model_peak_predictions}
\end{figure*}

Below several tens of TeV, the contribution of the Geminga supernova
remnant to the local cosmic-ray density gradient is relatively
significant, while parallel diffusion along the IBEX magnetic-field
direction causes the cosmic-ray flux to be directed toward the local
magnetic-field direction. The projection of the Geminga supernova
remnant in right ascension is approximately $82.5^\circ$, whereas the
projection of the local interstellar magnetic field measured by IBEX
in right ascension is approximately $48.5^\circ$. Under their combined
effects, the model predicts that the direction of maximum relative
intensity lies at approximately $60^\circ$--$70^\circ$, which is
basically consistent with the low-energy peak right ascensions measured
by the three experiments, as shown in
Fig.~\ref{fig:model_peak_predictions}.

As the energy increases, the nearby-source component associated with
the Geminga supernova remnant gradually weakens after approaching its
cutoff rigidity, while the relative contribution of the background
source component increases. In the phase-flip region, the anisotropy
vectors produced by the nearby source and the background sources point
in different directions and can partially cancel each other, causing
$A_{\max}$ to decrease and $\alpha_{\max}$ to flip rapidly. At high
energies, the background source component gradually becomes dominant,
and the peak direction evolves toward $250^\circ$--$300^\circ$.

\section{Conclusions and Outlook}

In this work, we perform a unified multi-harmonic fit to the one-dimensional relative-intensity distributions from ARGO-YBJ, Tibet~AS$\gamma$, and HAWC, and use the peak amplitude $A_{\max}$ and peak right ascension $\alpha_{\max}$ to summarize the main characteristics of each energy bin. The main conclusions are as follows.

First, in the common energy range of 11--22~TeV covered by the three experiments, the published first-harmonic phases have a range of $22.6^\circ$, whereas the analysis using $\alpha_{\max}$ reduces this range to $7.5^\circ$. This result shows that, after unifying the definition and extraction procedure of the observable characterizing the dominant cosmic-ray excess direction, the consistency among the three experiments is significantly improved.

Second, the values of $\alpha_{\max}$ from the three experiments form a trajectory that evolves continuously with energy. Below several tens of TeV, they are concentrated within $60^\circ$--$70^\circ$, between the right-ascension projections of the direction toward the Geminga supernova remnant and the direction of the IBEX local interstellar magnetic field. At high energies, they are located mainly within $250^\circ$--$300^\circ$, while $A_{\max}$ increases overall. In the phase-flip region, $A_{\max}$ exhibits a local minimum, and $\alpha_{\max}$ changes rapidly.

Third, the spatially dependent propagation model combines a Geminga-related instantaneous source with anisotropic diffusion along the direction of the local interstellar magnetic field. In the two tests with $\epsilon=0.10$ and $0.15$, while keeping the remaining parameters unchanged, the model describes the $P$, He, C, O, and Fe energy spectra. After applying the same right-ascension projection and maximum-based definitions to the model anisotropy, the model also reproduces the overall energy evolution of $A_{\max}$ and $\alpha_{\max}$. This provides a self-consistent physical explanation for large-scale anisotropy in the TeV to tens-of-TeV energy range.

In the future, this interpretation can be tested at both the data and observational levels. On the data side, more experiments should publish energy-binned one-dimensional relative-intensity distributions, or directly provide the peak amplitude and peak right ascension, so that they can be extracted using a unified procedure. Different analyses should adopt a consistent treatment of declination coverage and distinguish the contributions of non-dipolar structures such as the \textit{tail-in} and \textit{loss-cone}. On the observational side, the phase-flip region from tens to hundreds of TeV is the most critical and requires larger event statistics and higher energy resolution. If independent experiments confirm that $A_{\max}$ and $\alpha_{\max}$ exhibit energy evolution consistent with that found in this work, it will provide a stringent test of the joint interpretation involving a nearby Geminga source and the local interstellar magnetic field.

\begin{acknowledgments}
This work was financially supported by the National Natural Science Foundation of China (Grants No. 12373105,12333006, 12275279,  and 12320101005), the National Key R\&D Program of China (Grant No. 2025SKA0110103),  and the Hebei Province Graduate Student Innovation Project (Grant No. CXZZBS2026083).
\end{acknowledgments}

\appendix
\onecolumngrid

\section{}
\label{app:hawc_validation}

The HAWC one-dimensional relative-intensity distributions used in this work were extracted from eight published two-dimensional relative-intensity sky maps, corresponding to energies of 2.0, 3.0, 4.4, 6.8, 11.2, 18.6, 30.3, and 72.8~TeV \cite{Abeysekara2018HAWC}. To represent this type of elliptical sky map, a Mollweide projection was adopted, and the image positions of the two marked local interstellar magnetic-field directions were used as registration reference points. Subsequently, a regular $2^{\circ}\times2^{\circ}$ grid was constructed on each panel, with grid centers at right ascensions of $\alpha=1^{\circ},3^{\circ},\dots,359^{\circ}$ and declinations of $\delta=-30^{\circ},-28^{\circ},\dots,68^{\circ}$; an intensity-calibration table was constructed from the common color scale shared by all energy bins, and the color scale was mapped onto the interval $\delta I\in[-30,30]\times10^{-4}$. For each mapped grid point, the colors within the surrounding $7\times7$-pixel region were extracted and matched using the RGB-space distance
\[
d_{\mathrm{RGB}}
=
\sqrt{
(R-R_{\mathrm{c}})^2+
(G-G_{\mathrm{c}})^2+
(B-B_{\mathrm{c}})^2
},
\]
where $(R,G,B)$ denotes the color of a sky-map pixel and $(R_{\mathrm{c}},G_{\mathrm{c}},B_{\mathrm{c}})$ denotes a candidate color on the color scale. Pixels satisfying $d_{\mathrm{RGB}}\leq45$ were directly color-matched, and the median of the corresponding $\delta I$ values was assigned to the grid point. The regions covered by the black magnetic-field markers were first removed and subsequently filled by interpolation using the values of the nearest valid grid points, whereas extended no-data regions remained masked and excluded. The relative intensity was defined as $I=1+\delta I$.

\begin{figure}[!htbp]
    \centering
    \includegraphics[width=0.7\textwidth]{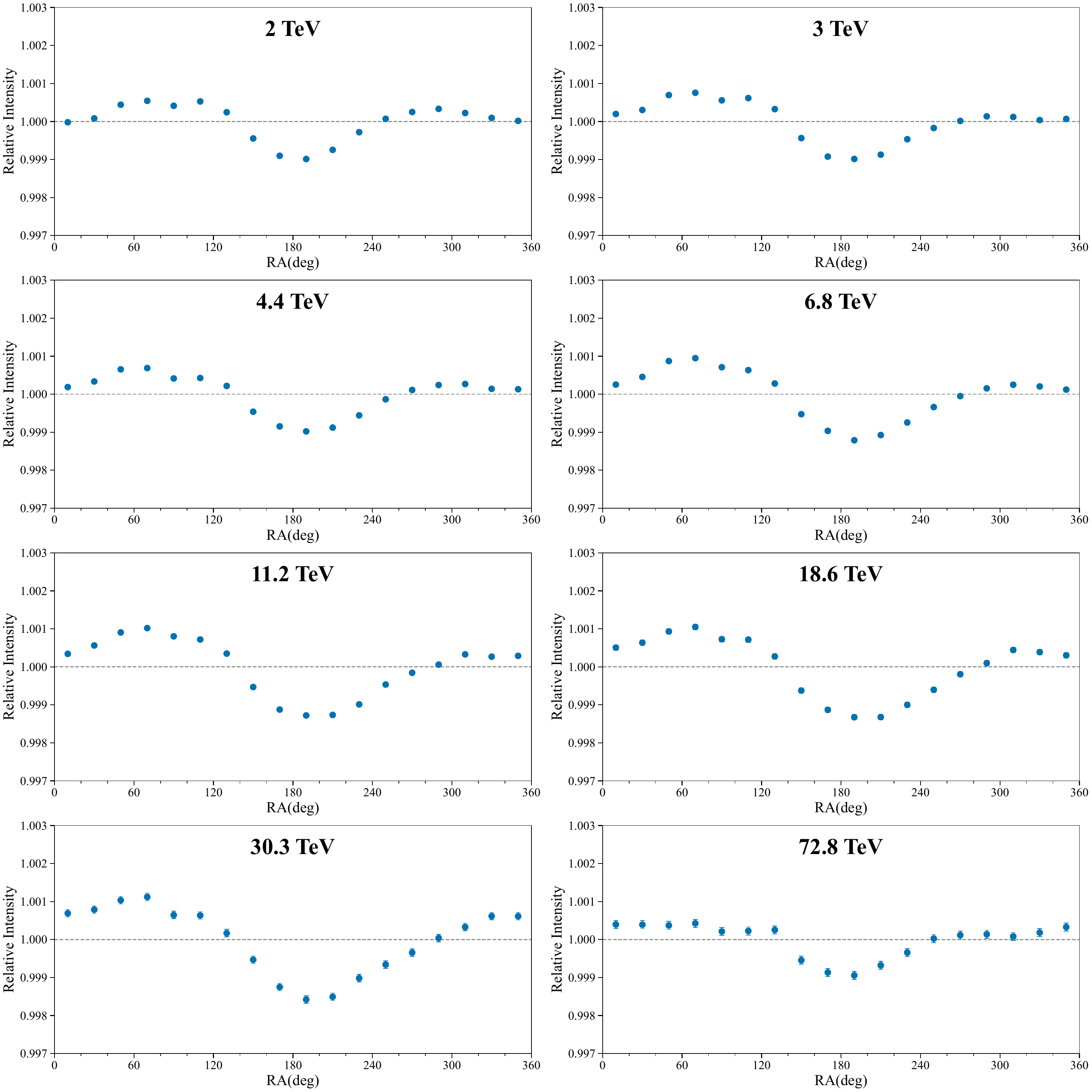}
    \caption{
        One-dimensional relative-intensity distributions extracted from the HAWC sky maps for the eight energy bins from 2 to 72.8~TeV~\cite{Abeysekara2018HAWC}. The horizontal dashed lines indicate a relative intensity of unity.
    }
    \label{fig:hawc_validation}
\end{figure}

To construct the one-dimensional right-ascension distribution, all valid $2^{\circ}\times2^{\circ}$ grid cells within each $2^{\circ}$ right-ascension interval were first averaged along the declination direction. Every 10 consecutive $2^{\circ}$ right-ascension data points were then grouped and averaged into a single $20^{\circ}$ right-ascension bin, yielding 18 data points centered at $\alpha=10^{\circ},30^{\circ},\dots,350^{\circ}$. Assuming that the total number of events $N$ in each energy bin was uniformly distributed among the 18 bins, a common Poisson statistical uncertainty,
\[
\sigma_I=\sqrt{\frac{18}{N}},
\]
was assigned to all data points in that energy bin. The resulting one-dimensional relative-intensity distributions are shown in Fig.~\ref{fig:hawc_validation}.

\section{}
\label{app:argo22_harmonic_fits}

\begin{figure}[!htbp]
    \centering
    \includegraphics[width=0.75\textwidth]{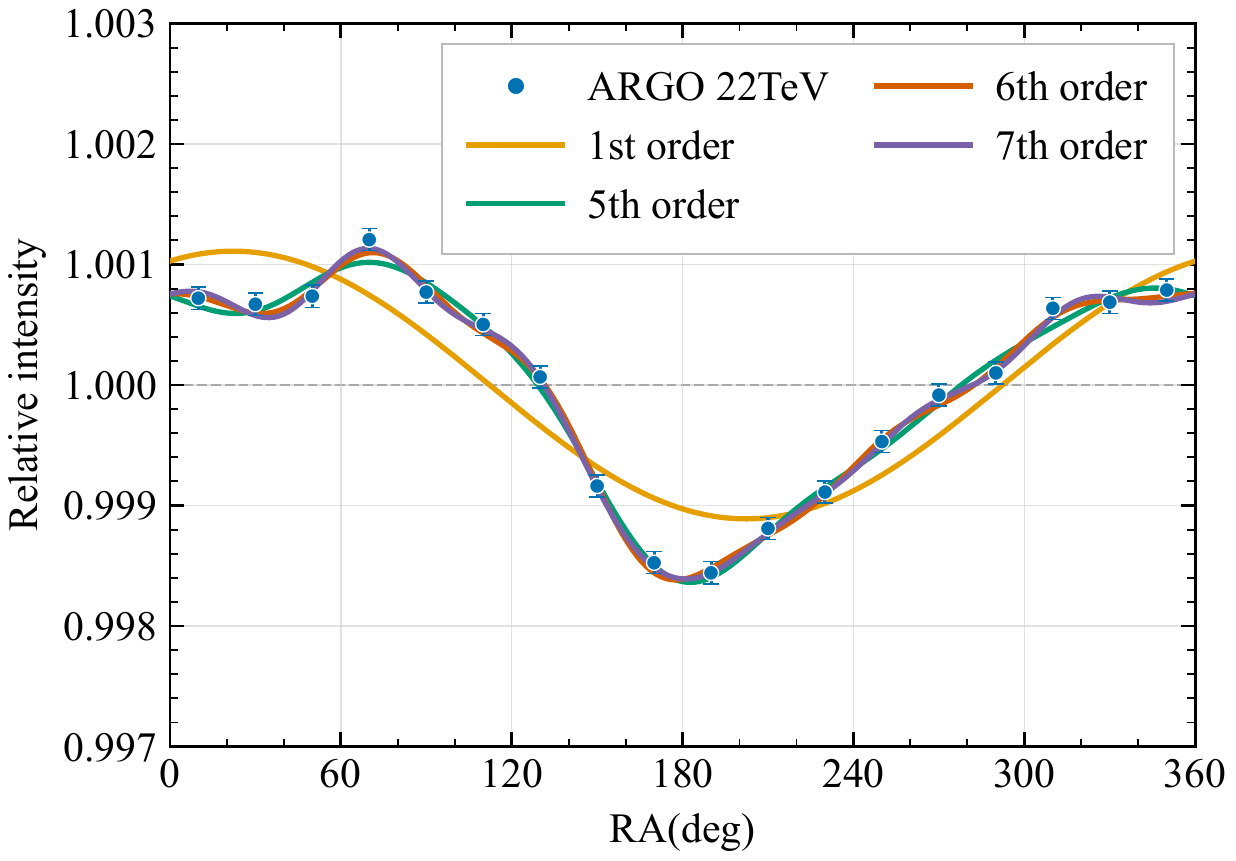}
    \caption{The one-dimensional relative-intensity distribution for the 22~TeV ARGO-YBJ energy bin~\cite{Bartoli2018ARGO}. The yellow, green, red, and purple curves represent the first-, fifth-, sixth-, and seventh-order harmonic fits, respectively. The error bars indicate the statistical uncertainties adopted in this work, and the horizontal dashed line indicates a relative intensity of unity.}
    
    \label{fig:argo22_harmonic_fits}
\end{figure}

For the 22~TeV ARGO-YBJ relative-intensity distribution, the peak right ascensions obtained from the fifth-, sixth-, and seventh-order harmonic fits are $69.9^\circ$, $71.3^\circ$, and $69.7^\circ$, respectively. The differences between adjacent higher-order results are all within the adopted $2^\circ$ convergence threshold; the sixth- and seventh-order peak right ascensions differ only slightly from the fifth-order result, whereas the peak right ascensions obtained from the fourth- and fifth-order fits differ by $4.1^\circ$. Therefore, the fifth-order reconstruction is selected as the lowest-order model satisfying the convergence criterion. Figure~\ref{fig:argo22_harmonic_fits} shows the first-order fit together with the fifth-, sixth-, and seventh-order reconstructions.

\twocolumngrid
\bibliography{references}

\end{document}